\documentclass[12pt, a4paper]{article}
\newcommand{\D}{\mbox{\rm d}}
\newcommand{\Tr}{\mbox{\rm Tr}}
\begin{document}
\title{Positive-definite states of a Klein-Gordon type
particle\footnote{Physics Letters A {\bf 372}, 4180 (2008)}}
\author{A.A. Semenov$^{1,2}$, C.V. Usenko$^{3}$, and B.I. Lev$^{1,2}$
\\ \quad \\
{\footnotesize $^1$Bogolyubov Institute for Theoretical Physics,}\\
{\footnotesize National Academy of Sciences of
Ukraine,}\\{\footnotesize Vul. Metrologichna 14-b,
UA-03680 Kiev, Ukraine}\\
{\footnotesize $^{2}$Institute of Physics,
National Academy of Sciences of Ukraine,}\\
{\footnotesize Prospect Nauky 46,
UA-03028 Kiev, Ukraine}\\
{\footnotesize $^3$Physics Department, Taras Shevchenko National University of Kiev,}\\
{\footnotesize Prospect Glushkova 2,
UA-03127 Kiev, Ukraine}\\
{\footnotesize E-mail: sem@iop.kiev.ua}}

\maketitle

\begin{abstract}
A possible way for the consistent probability interpretation of the
Klein-Gordon equation is proposed. It is assumed that some states of
a scalar charged particle cannot be physically realized. The rest of
quantum states are proven to have positive-definite probability
distributions.
\end{abstract}

PACS: 03.65.Pm, 03.65.Yz, 12.90.+b

\section{Introduction}
\label{int}

The problem of physical interpretation of the Klein-Gordon equation
has a very long history. In the early days of quantum mechanics,
Schr\"odinger \cite{Schroedinger}, Klein \cite{Klein}, Gordon
\cite{Gordon}, and Fock \cite{Fock} considered it as a relativistic
counterpart of the Schr\"odinger wave equation. However, it was
impossible to ascribe a consistent probability interpretation to
solutions of this equation due to negative values of the
corresponding distribution functions, see e.g. \cite{Gerlach1}. The
four-component quantum relativistic wave equation \cite{Dirac},
proposed later by Dirac, is free of this problem. This is why the
Dirac equation was considered as the unique possible way for the
consistent formulation of relativistic quantum mechanics.

Nevertheless, as it has been demonstrated later by Pauli and
Weisskopf \cite{Pauli}, quantized Klein-Gordon field consistently
describes spinless charged particles, $\pi^{\pm}$ mesons.
Probability distributions for the total field energy and the field
energy in a space region are positive definite. In the case of free
particles this feature is also proper to the probability
distribution of momentum. Procedures of measuring these observables
are well-defined for Klein-Gordon type particles as well as for
Dirac $1/2$-spin particles.

At the same time, the measurement of the position is a special
problem in the relativistic quantum theory. Indeed, according to the
von Neumann reduction postulate \cite{NeumannBook}, a result of the
corresponding procedure is that the particle occurs to be in an
eigenstate of the position operator. However, as it follows from the
Hegerfeldt theorem \cite{Hegerfeldt}, particles cannot be localized
in the finite space region. Consequently, a localized state cannot
appear as a result of any measurement procedure. Moreover, such a
state would be a superposition of two states with different signs of
charge that is prohibited by the fundamental charge superselection
rule \cite{Superselection}.

From the other hand, the von Neumann measurement is only a special
type of the measurement procedures. More general situation occurs
when the resulting state differs from the eigenstate of the measured
observables. It happens, for example, in the well-studied process of
photodetection. In this case, the detection of photons means their
absorption. The resulting state consists fewer photons -- generally
it differs from the Fock numberstate, see e.g. \cite{Gardiner}.
Similarly one may assume, without referring to any special physical
procedure, that in the case of the measurement of the position,
resulting states differ from the prohibited eigenstates of the
position operator. Consequently, the impossibility of localization
does not lead to the impossibility of measuring the position, as it
is frequently assumed.

As it was mentioned above, Dirac particles always have
positive-definite position probability distribution \cite{Bracken}.
In this case there are no conceptual problems for the measurement of
the position. However such a problem arises for Klein-Gordon
particles -- the corresponding probability distribution may have
negative values. This problem is addressed in the present
contribution.

Newton and Wigner introduced the position operator different from
the standard one \cite{Newton}. Among other, the probability
distribution for this observable has non-negative values even for
the case of Klein-Gordon particles. However, this approach meets at
least two serious problems -- absence of the Lorentz-invariance and
difficulties in formulation of the operational procedure for
measuring such an observable.

A way for resolving this problem has been proposed by Gerlach,
Gromes, and Petzold \cite{Gerlach}. It has been shown that the
current density of the Klein-Gordon field can be consistently
redefined in such a way that the corresponding time component,
charge density, is positive definite. Similarly, Mostafazadeh has
proposed to redefine the scalar product for the solutions of the
Klein-Gordon equation \cite{Mostafazadeh}. In the framework of this
approach one also gets positive-definite probability distributions
for the position.

In this contribution we propose an alternative solution of this old
problem that does not require the redefinition of the standard
scalar product. Our approach is based on the standard requirement
that quantum states of a system can be described only by
positive-definite density operator. It will be demonstrated that
among the pure states of a Klein-Gordon type particle only
eigenstates of the Hamiltonian obey this condition. Hence, all
non-stationary positive-definite states are mixed ones.

\section{Effective density operator}
\label{edo}

Let us start from the well-known Feshbach-Villars form of the
Klein-Gordon equation \cite{Feshbach}
\begin{equation}
i\hbar\frac{\partial}{\partial
t}\left|\Psi\right\rangle=\hat{H}\left|\Psi\right\rangle,\label{eq1}
\end{equation}
where the Hamiltonian
\begin{equation}
\hat{H}=\left(\tau_3+i\tau_2\right)\frac{\hat{p}^2}{2m} +\tau_3 mc^2
\label{eq2}
\end{equation}
is $2\times 2$ operator-valued matrix and $\tau_i$ are the Pauli
matrices. These matrices describe an internal degree of freedom,
so-called charge variable, associated with the sign of frequency
that is sign of the charge in the quantum field theory. Utilizing
the generalized unitary transformation defined by the
operator-valued matrix
\begin{equation}
\hat{S}=\frac{1}{2\sqrt{mc^2\hat{E}}}\left[\left(\hat{E}+mc^2\right)
+\left(\hat{E}-mc^2\right)\tau_1\right] \label{eq3}
\end{equation}
and the inverted one
\begin{equation}
\hat{S}^{-1}=\frac{1}{2\sqrt{mc^2\hat{E}}}\left[\left(\hat{E}+mc^2\right)
-\left(\hat{E}-mc^2\right)\tau_1\right], \label{eq4}
\end{equation}
where
\begin{equation}
\hat{E}=\sqrt{m^2c^4+c^2\hat{ p}^2}, \label{eq5}
\end{equation}
the Hamiltonian (\ref{eq2}) can be diagonalized to the form
\begin{equation}
\hat{H}^{\mathrm{FV}}=\hat{S}\hat{H}\hat{S}^{-1}=\tau_3\,\hat{E}.
\label{eq6}
\end{equation}
The corresponding representation is referred to as the
Feshbach-Villars (FV) represen\-tation. The same rule,
\begin{equation}
\hat{A}^{\mathrm{FV}}=\hat{S}\hat{A}\hat{S}^{-1},\label{eq7}
\end{equation}
is applied for transformation of an arbitrary observable, $\hat{A}$,
in the FV representation.

Consider the observable, $\hat{A}$, that is a function of the
position, $\hat{q}$, and momentum, $\hat{p}$; moreover, we suppose
that this observable does not depend on the matrices $\tau_i$,
\begin{equation}
\hat{A}=f\left(\hat{p},\hat{ q}\right).\label{eq8}
\end{equation}
The application of eqs.~(\ref{eq3}), (\ref{eq4}), (\ref{eq7}) yields
\begin{equation}
\hat{A}^{\mathrm{FV}}=\left(\mathcal{E}+\tau_1\mathcal{X}\right)\hat{A}.
\label{eq9}
\end{equation}
In the last equations $\mathcal{E}$ and $\mathcal{X}$ are
superoperators, which take the operator $\hat{A}$ to the operators
$\mathcal{E}\hat{A}$ and $\mathcal{X}\hat{A}$, respectively, as
\begin{eqnarray}
\left\langle p_2 \right|\mathcal{E}\hat{A}\left|p_1\right\rangle
=\varepsilon\left(p_2,p_1\right)\left\langle p_2
\right|\hat{A}\left|p_1\right\rangle,\label{eq10}\\ \left\langle p_2
\right|\mathcal{X}\hat{A}\left|p_1\right\rangle
=\chi\left(p_2,p_1\right)\left\langle p_2
\right|\hat{A}\left|p_1\right\rangle,\label{eq11}
\end{eqnarray}
where
\begin{eqnarray}
\varepsilon\left(p_2,p_1\right)=\frac{{E(p_2)+E(p_1)}}{{2\sqrt
{E(p_2)E(p_1)}}},\label{eq12} \\
\chi\left(p_2,p_1\right)=\frac{{E(p_2)-E(p_1)}}{{2\sqrt{E(p_2)E(p_1)}}},
\label{eq13}
\end{eqnarray}
and $\left|p\right\rangle$, $E\left(p\right)$ are eigenstates and
eigenvalues of the operator (\ref{eq5}).

As an example, consider the standard position operator, $\hat{q}$.
In this case,
\begin{equation}
\mathcal{E}\hat{q}=\hat{\xi},
\end{equation}
where $\hat{\xi}$ is the Newton-Wigner position operator
\cite{Newton} in the FV representation,
\begin{equation}
\hat{\xi}=i\hbar\frac{\partial}{\partial p},\label{NW}
\end{equation}
that is the even part of the standard position operator
\cite{Feshbach}. Similarly,
\begin{equation}
\tau_1\mathcal{X}\hat{q}=-i\hbar\frac{c^2 \hat{p}}{2E^2
(\hat{p})}\tau_1
\end{equation}
is the odd part of the position operator. We recall that the even
part of an observable is a diagonal matrix in the FV representation
and the odd part is an off-diagonal matrix.

Any state of a scalar charged particle satisfies the charge
superselection rule \cite{Superselection}, i.e. it does not include
interference terms between particle and antiparticle. Moreover,
without loss of generality one can consider only the positive sign
of the charge. This means that the expected value of the observable
(\ref{eq8}), (\ref{eq9}) can be written as
\begin{equation}
\bar{A}=\Tr\left(\hat{\varrho}\,\mathcal{E}\!\hat{A}\right),
\label{eq14}
\end{equation}
where $\hat{\varrho}$ is the density operator in the FV
representation. Specifically, the even part of $\hat{q}^2$ is given
by
\begin{equation}
\mathcal{E}\hat{q}^2=\hat{\xi}^2-\left[\hbar\frac{c^2 \hat{p}}{2E^2
(\hat{p})}\right]^2.\label{SecMomPos}
\end{equation}
Hence, the variance of the position,
\begin{equation}
\left\langle\Delta q^2\right\rangle=\left\langle\Delta \xi
^2\right\rangle-\left\langle\left[\frac{\hbar c^2\hat{
p}}{2E^2\left(\hat{p}\right)}\right]^2 \right\rangle,\label{PosVar}
\end{equation}
may be negative \cite{Lev1} that is a consequence of the fact that
the probability distribution is sign indefinite.

Eq.~(\ref{eq14}) can be considered as a scalar product of two
elements, $\hat{\varrho}$ and $\mathcal{E}\!\hat{A}$, in the
operator space. The action of the Hermitian-conjugated superoperator
$\mathcal{E}$ can be transferred from $\hat{A}$ to $\hat{\varrho}$
in eq.~(\ref{eq14}), i.e.
$\Tr\left(\hat{\varrho}\,\mathcal{E}\!\hat{A}\right)=
\Tr\left(\mathcal{E}\!\hat{\varrho}\,\hat{A}\right)$. This fact can
be simply proved by using Eqs.~(\ref{eq12}), (\ref{eq13}). Further
introducing the operator
\begin{equation}
\hat{\varrho}_{{}_{\mathcal{E}}}=\mathcal{E}\hat{\varrho},\label{eq15}
\end{equation}
which we refer to as the effective density operator,
eq.~(\ref{eq14}) is rewritten,
\begin{equation}
\bar{A}=\Tr\left(\hat{\varrho}_{{}_\mathcal{E}}\hat{A}\right).
\label{eq16}
\end{equation}
Similarly,  the probability distribution to get the value $A$ is
\begin{equation}
P\left(A\right)=\Tr\Big(\hat{\varrho}_{{}_\mathcal{E}}\left|A\right\rangle
\left\langle A\right|\Big), \label{ProbDist}
\end{equation}
where $A$ and $\left|A\right\rangle$ are eigenvalues and
eigenstates, respectively, of the operator $\hat{A}$.

As it follows from eq.~(\ref{eq15}), the superoperator $\mathcal{E}$
takes each density operator, $\hat{\varrho}$, to the effective
density operator, $\hat{\varrho}_{{}_{\mathcal{E}}}$. Since
\begin{eqnarray}
\varepsilon\left(p_2,p_1\right)=\varepsilon\left(p_1,p_2\right)>1&\,
\textrm{for}&\, p_2\neq p_1,\label{Prop1}\\
\varepsilon\left(p,p\right)=1,\label{Prop2}
\end{eqnarray}
there exist sign-indefinite effective density operators,
$\hat{\varrho}_{{}_{\mathcal{E}}}$, assigned to the
positive-definite density operators, $\hat{\varrho}$. Consequently,
the probability distribution (\ref{ProbDist}) may have negative
values even for some positive-definite density operators,
$\hat{\varrho}$. This is the reason of problems in consistent
probability interpretation of the Klein-Gordon equation.

The above consideration enables us to formulate the following
assumption. Let us suppose that the states of a Klein-Gordon type
particle are described by the positive-definite effective density
operators, $\hat{\varrho}_{{}_\mathcal{E}}$. Inverting the map
(\ref{eq15}), we get the following condition for the density
operators:
\begin{equation}
\hat{\varrho}=\mathcal{E}^{-1}\hat{\varrho}_{{}_\mathcal{E}},
\label{eq17}
\end{equation}
where $\hat{\varrho}_{{}_\mathcal{E}}$ is a positive-definite
operator for which
$\Tr\left(\hat{\varrho}_{{}_\mathcal{E}}\right)=1$,
$\mathcal{E}^{-1}$ is the superoperator inverted to $\mathcal{E}$
and defined by the rule
\begin{equation}
\left\langle p_2
\right|\mathcal{E}^{-1}\hat{\varrho}_{{}_\mathcal{E}}\left|p_1\right\rangle
=\varepsilon^{-1}\left(p_2,p_1\right)\left\langle p_2
\right|\hat{\varrho}_{{}_\mathcal{E}}\left|p_1\right\rangle.\label{eq18}
\end{equation}
Taking into account that
\begin{eqnarray}
\varepsilon^{-1}\left(p_2,p_1\right)=\varepsilon^{-1}\left(p_1,p_2\right)<1&\,
\textrm{for}&\, p_2\neq p_1,\label{eq19}\\
\varepsilon^{-1}\left(p,p\right)=1,\label{eq20}
\end{eqnarray}
we conclude that under the above assumption the states of a
Klein-Gordon type particle are mostly mixed. Only eigenstates of the
Hamiltonian are pure and obey the condition (\ref{eq17}).

\section{Observables}
\label{Obs}

The above consideration deals with the observables that are
arbitrary functions of position and momentum. However, some
observables do not belong to this class. For example, the
Hamiltonian (\ref{eq2}) is a function of position, momentum, and
charge variable. At the same time, the expected value of the
Hamiltonian is determined only by diagonal matrix elements of the
density operator. As it follows from eqs.~(\ref{Prop2}),
(\ref{eq20}),
\begin{equation}
\left\langle p
\right|\hat{\varrho}_{{}_\mathcal{E}}\left|p\right\rangle=\left\langle
p \right|\hat{\varrho}\left|p\right\rangle.
\end{equation}
Hence, the expected value of the Hamiltonian is given by
\begin{equation}
\bar{H}=\Tr\left(\hat{\varrho}_{{}_\mathcal{E}}\hat{E}\right),
\label{Mean Energy}
\end{equation}
where $\hat{E}$ is defined by eq.~(\ref{eq5}). This result adopts
the rule (\ref{eq8}) for the evaluation of the expected value of the
Hamiltonian.

Another example is the velocity operator,
\begin{equation}
\hat{v}=\frac{1}{i\hbar}\left(\hat{q}\hat{H}-\hat{H}\hat{q}\right),
\end{equation}
which is a function of the position, $\hat{q}$, and the Hamiltonian,
$\hat{H}$. More generally, let us consider arbitrary functions of
position, momentum, and energy, i.e.
\begin{equation}
\hat{B}=F\left(\hat{ p},\hat{q}, \hat{H}\right).\label{GenObs}
\end{equation}
In the FV representation this observable has the form
\begin{equation}
\hat{B}^{\mathrm{FV}}=\left(\mathcal{E}-\tau_1\mathcal{X}\right)F\left(\hat{
p},\hat{\xi}, \tau_3\hat{E}\right),
\end{equation}
where $\hat{\xi}$ is the Newton-Wigner position operator,
eq.~(\ref{NW}). It now follows that the expected value of this
observable is given by
\begin{equation}
\bar{B}=\Tr\left[\hat{\varrho}\,\mathcal{E}\!F\!\left(\hat{
p},\hat{\xi}, \hat{E}\right)\right],
\end{equation}
where we suppose the positive sign of the charge. In terms of the
effective density operator it is rewritten as
\begin{equation}
\bar{B}=\Tr\left[\hat{\varrho}_{{}_\mathcal{E}}\,F\!\left(\hat{
p},\hat{\xi}, \hat{E}\right)\right].
\end{equation}
Arguing as above, we conclude that for the observable (\ref{GenObs})
one can also apply the effective density operator (\ref{eq15}) and
require its positivity.

\section{Relativistic phase-space representation}
\label{RelPhS}

An example of the positive-definite states is a natural
generalization of the coherent state $\left|\alpha\right\rangle$
\begin{equation}
\hat{\varrho}_{\alpha}=\mathcal{E}^{-1}\left|\alpha\right\rangle
\left\langle\alpha\right|,\label{eq21}
\end{equation}
where $\left|\alpha\right\rangle$ is an eigenstate of the operator
\begin{equation}
\hat{a}=\frac{1}{\sqrt{2}}\left(\frac{\hat{\xi}}{\lambda}+
i\frac{\lambda}{\hbar}\hat{p}\right),
\end{equation}
$\hat{\xi}$ is the Newton-Wigner position operator (\ref{NW}),
$\lambda$ is a scaling length. Contrary to the non-relativistic
case, for a review see e.g. \cite{Klauder}, the coherent state
(\ref{eq21}) is not pure. However, it manifests the properties
essential for the usual coherent states. Particularly, the coherent
states (\ref{eq21}) minimize the Heisenberg uncertainty relation,
\begin{equation}
\left\langle\Delta q^2\right\rangle\left\langle\Delta
p^2\right\rangle=\frac{\hbar^2}{4}.
\end{equation}
At the same time, the pure relativistic coherent states of a scalar
charged particle, see e.g. \cite{Lev1}, \cite{Malkin},
\cite{Prugovecki}, \cite{Ali} and references therein, are not
positive definite and do not obey the condition (\ref{eq17}).

Similarly to \cite{Glauber}, we present the $s$-parameterized
phase-space distribution of a scalar charged particle as
\begin{equation}
P\left(\alpha;s\right)=\frac{1}{\pi^2}
\int\limits_{-\infty}^{+\infty}\D^2\beta\,
\Tr\left[\hat{\varrho}_{{}_\mathcal{E}}\,e^{(\hat{a}^{\dag}-\alpha^*)\beta-
(\hat{a}-\alpha)\beta^*+s\frac{|\beta|^2}{2}}\right].
\label{SParametrDistr}
\end{equation}
For $s=0$ eq.~(\ref{SParametrDistr}) defines the relativistic Wigner
function considered in \cite{Lev2}, for $s=1$ -- the relativistic
Glauber-Sudarshan distribution, for $s=-1$ -- the relativistic
Husimi-Kano distribution. The density operator $\hat\varrho$ can be
expressed in terms of the Glauber-Sudarshan distribution
\cite{GlauberSudarshan},
\begin{equation}
\hat{\varrho}=\int\limits_{-\infty}^{+\infty}\D^2\alpha
P\left(\alpha;1\right) \hat{\varrho}_{\alpha},
\end{equation}
where $\hat{\varrho}_{\alpha}$ is given by eq.~(\ref{eq21}).
Likewise, the Husimi-Kano distribution \cite{HusimiKano}, can be
expressed in terms of the density operator $\hat\varrho$,
\begin{equation}
P\left(\alpha;-1\right)=\frac{1}{\pi}\Tr\left(\hat{\varrho}\hat{\varrho}_{\alpha}\right).
\end{equation}
The phase-space distribution (\ref{SParametrDistr}) includes the
complete information about the density operator and can be used for
the characterization of the quantum states of a Klein-Gordon type
particle.

\section{Conclusions}
\label{Concl}

We have demonstrated that the reason of problems in consistent
probability interpretation of the Klein-Gordon equation is that a
lot of quantum states are described by sign-indefinite density
operators. Particularly, the most of pure states are sign
indefinite. From the other hand, there exist the states of a
Klein-Gordon type particle that are characterized by
positive-definite density operators. We suppose that only such
states have physical realization. This assumption is in a good
accordance with the basic principles of quantum physics.

The eigenstates of the Hamiltonian are positive definite. Since such
states are stationary, the unitary evolution does not result in the
appearance of negative-definite states. The positive-definite
non-stationary states of a Klein-Gordon type particle are mixed. The
unitary evolution cannot be characterized by decreasing the entropy,
i.e., it cannot purify the states. In this reason such states are
positive-definite for any moment of time.

In the outlook we address the following problems. First of all, the
question about the Lorentz invariance of the proposed approach is
still open. In other words, the question is whether the state, which
is positive definite in a given reference frame, is positive
definite in other reference frames? Another problem is in the
generalization of the proposed approach to the case of quantized
Klein-Gordon field. We believe that resolving these problems results
in the progress of the scalar charged particles theory.


\begin{thebibliography}{99}
\bibitem{Schroedinger} E. Schr\"odinger, Ann. Phys. {\bf 81},
109 (1926) .
\bibitem{Klein} O. Klein, Z. Phys. {\bf 37}, 895 (1926).
\bibitem{Gordon} W. Gordon, Z. Phys. {\bf 40}, 117 (1926);
{\bf 40}, 121.
\bibitem{Fock} V.A. Fock, Z. Phys. {\bf 38}, 242 (1926); {\bf 39},
226.
\bibitem{Gerlach1} B. Gerlach, D. Gromes, J. Petzold, and P.
Rosenthal, Z. Phys. {\bf 202}, 401 (1967).
\bibitem{Dirac} P.A.M. Dirac Proc. R. Soc. {\bf 117}, 610 (1928).
\bibitem{Pauli} W. Pauli and V. Weisskopf, Helv. Phys. Acta. {\bf
7}, 709 (1934).
\bibitem{NeumannBook} J. von Neumann, Mathematical Foundations of
Quantum Mechanics, Princeton Univ. Press, Princeton, 1955.
\bibitem{Hegerfeldt} G.C. Hegerfeldt, Phys. Rev. D. {\bf 10}, 3320
(1974).
\bibitem{Superselection} G.C. Wick, A.S. Wightman and E.P. Wigner
Phys. Rev. {\bf 88}, 101 (1952); E.P. Wigner Z. Phys. {\bf 133}, 101
(1952).
\bibitem{Gardiner}C. W. Gardiner, P. Zoller, {\em Quantum noise},
(Berlin: Springer, 2000).
\bibitem{Bracken} A.J. Bracken and G.F. Melloy, J. Phys. A {\bf 32}, 6127 (1999).
\bibitem{Newton} T.D. Newton and E.P. Wigner, Rev. Mod. Phys. {\bf 21}, 400 (1949).
\bibitem{Gerlach} B. Gerlach, D. Gromes, and J. Petzold, Z. Phys. {\bf
204}, 1 (1967).
\bibitem{Mostafazadeh} A. Mostafazadeh, Ann. Phys. NY {\bf 309}, 1 (2004).
\bibitem{Feshbach} H. Feshbach and F. Villars, Rev. Mod. Phys. {\bf
30}, 24 (1958).
\bibitem{Lev1} B.I. Lev, A.A. Semenov, C.V. Usenko, and J.R. Klauder Phys. Rev. A
{\bf 66}, 022115 (2002).
\bibitem{Klauder} J.R. Klauder and B.-S. Skagerstam, {\em Coherent
States, Applications in Physics and Mathematical Physics}
(Singapore: World Scientific, 1982).
\bibitem{Malkin}I. A. Malkin and V. I. Man'ko, Sov. Phys.--JETP
{\bf 28}, 527 (1969).
\bibitem{Prugovecki} E. Prugovecki, {\em Stochastic Quantum Mechanics and Quantum
Spacetime}, (Dordrecht: Reidel, 1984).
\bibitem{Ali} S.T. Ali, Rivista del Nuovo Cimento {\bf 8}, 1 (1985).
\bibitem{Glauber} K.E. Cahil and R.J. Glauber, Phys. Rev. {\bf 177}
1857, 1882 (1969).
\bibitem{GlauberSudarshan} R.J. Glauber, Phys. Rev.
Lett. {\bf 10}, 84 (1963); Phys. Rev. {\bf 6}, 2766 (1963); E.C.G.
Sudarshan, Phys. Rev. Lett. {\bf 10}, 277 (1963).
\bibitem{HusimiKano} K. Husimi, Proñ. Phys. Math. Soc. Japan {\bf 22}, 264 (1940);
Y. Kano, J. Math. Phys. {\bf 6}, 1913 (1965).
\bibitem{Lev2} B.I. Lev, A.A. Semenov, and C.V. Usenko, J. Phys. A {\bf
34}, 4323 (2001); J. Rus. Las. Res. {\bf 23}, 347 (2002).
\end{thebibliography}
\end{document}